\documentclass[fleqn,usenatbib]{mnras}

\usepackage{newtxtext,newtxmath}

\usepackage[T1]{fontenc}

\DeclareRobustCommand{\VAN}[3]{#2}
\let\VANthebibliography\thebibliography
\def\thebibliography{\DeclareRobustCommand{\VAN}[3]{##3}\VANthebibliography}

\usepackage{graphicx}	
\usepackage{amsmath}	

\title[Shape and scatter of IGM damping wings]{Investigating the characteristic shape and scatter of intergalactic damping wings during reionization}

\author[L. C. Keating et al.]{Laura C. Keating$^{1}$\thanks{laura.keating@ed.ac.uk},  Ewald Puchwein$^{2}$, James S. Bolton$^{3}$,  Martin G. Haehnelt$^{4}$ and Girish Kulkarni$^{5}$
\\
$^{1}$Institute for Astronomy, University of Edinburgh, Blackford Hill, Edinburgh, EH9 3HJ, UK\\
$^{2}$Leibniz-Institut f\"ur Astrophysik Potsdam, An der Sternwarte 16, 14482 Potsdam, Germany\\
$^{3}$School of Physics and Astronomy, University of Nottingham, University Park, Nottingham, NG7 2RD, UK\\
$^{4}$Kavli Institute for Cosmology and Institute of Astronomy, Madingley Road, Cambridge, CB3 0HA, UK\\
$^{5}$Tata Institute of Fundamental Research, Homi Bhabha Road, Mumbai 400005, India\\
}

\date{Accepted ---. Received ---; in original form ---}

\pubyear{2023}

\begin{document}
\label{firstpage}
\pagerange{\pageref{firstpage}--\pageref{lastpage}}
\maketitle

\begin{abstract}
Ly$\alpha$ damping wings in the spectra of bright objects at high redshift are a useful probe of the ionization state of the intergalactic medium during the reionization epoch. It has recently been noted that, despite the inhomogeneous nature of reionization, these damping wings have a characteristic shape which is a strong function of the volume-weighted average neutral hydrogen fraction of the intergalactic medium. We present here a closer examination of this finding using a simulation of patchy reionization from the Sherwood-Relics simulation suite. We show that the characteristic shape and scatter of the damping wings are determined by the average neutral hydrogen density along the line of sight, weighted by its contribution to the optical depth producing the damping wing. We find that there is a redshift dependence in the characteristic shape due to the expansion of the Universe. Finally, we show that it is possible to differentiate between the shapes of damping wings in galaxies and young (or faint) quasars at different points in the reionization history at large velocity offsets from the point where the transmission first reaches zero.
\end{abstract} 




\section{Introduction}

Lyman-$\alpha$ (Ly$\alpha$) scattering redward of rest-frame Ly$\alpha$ in the spectra of luminous objects is a clear indicator of the presence of neutral hydrogen. In the reionization epoch, even the neutral regions of the diffuse intergalactic medium (IGM) can cause such a ``damping wing'' \citep{miraldaescude1998}. Observations of damping wings in the spectra of $z > 7$ quasars indicate that the reionization of the Universe is in progress at these redshifts \citep{fan2022}. IGM damping wings have also been observed in the spectra of high-redshift gamma-ray burst afterglows \citep{hartoog2015} and galaxies \citep{umeda2023}. However, translating the strength of observed damping wings to a constraint on the ionization state of the IGM is non-trivial, in part due to the inhomogeneity of reionization \citep{mesinger2008, mcquinn2008}. The typical approach is to marginalise over the position of the quasar host halo within the large-scale morphology of the ionization field \citep{greig2017, davies2018}.

Recently, another approach was proposed by \citet{chen2023} using mock IGM damping wings generated from the CROC cosmological radiative transfer simulations of reionization \citep{gnedin2014}. In that work, the simulated damping wings were shifted along the wavelength axis, such that the IGM transmission $T$ along all lines of sight reached zero at the same point. Despite the spatially inhomogeneous distribution of neutral gas, these realigned damping wings were shown to have a characteristic shape, which was a strong function of the IGM volume-weighted average neutral fraction $\langle x_{\rm HI} \rangle_{\rm v}$. Applying the same technique to observations may simplify analyses of damping wings in the spectra of high-redshift objects, with \citet{chen2023} predicting that measurements of the volume-weighted average IGM neutral fraction of the order of $\Delta \langle x_{\rm HI} \rangle_{\rm v} \sim 0.1$ could be possible.

In this Letter, we investigate this characteristic shape of intergalactic damping wings in more detail using one of the Sherwood-Relics simulations of inhomogeneous reionization. We demonstrate why this characteristic shape arises. We also explain the origin and quantify the size of the scatter among the realigned transmission curves at a given volume-weighted average IGM neutral fraction. In Section \ref{sec:simulations}, we describe this simulation and outline how we produce the mock IGM damping wings. In Section \ref{sec:shape}, we discuss the origin of this characteristic shape and its scatter and, in Section \ref{sec:obs}, we make predictions for the observability of this signal in mock observations of damping wings in galaxies and quasars. Finally, in Section \ref{sec:conclusions}, we present our conclusions. We assume throughout this work that $\Omega_{\rm m} = 0.308$, $\Omega_{\Lambda} = 0.692$ and $h = 0.678$ \citep{planck2014}.

\section{Mock intergalactic damping wings}
\label{sec:simulations}

We analyse a simulation of inhomogeneous reionization from the Sherwood-Relics simulation suite\footnote{\url{https://www.nottingham.ac.uk/astronomy/sherwood-relics}} \citep{bolton2017, puchwein2023}. This simulation was performed with a modified version of \textsc{p-gadget-3} \citep{springel2005}. It has a box size of $40 \, h^{-1}$ cMpc and a gas particle mass of $M_{\rm gas} = 9.97 \times 10^{4} \, h^{-1} \, M_{\odot}$. Patchy reionization was treated with a novel hybrid technique described in detail in \citet{puchwein2023}. In brief, a cosmological hydrodynamic simulation was performed and then post-processed with the radiative transfer code \textsc{aton} \citep{aubert2008, aubert2010}. Maps of the spatially varying UV background from the radiative transfer simulation at finely spaced redshift intervals were used as input to a new cosmological hydrodynamic simulation, which then accounts for spatial fluctuations in the ionization state of the gas and its hydrodynamical response due to inhomogeneous reionization. 

Mock intergalactic damping wings were generated as described in \citet{keating2023}. Lines of sight were extracted in six orthogonal directions through the 100 most massive haloes in each snapshot, for a total of 600 sightlines with length 20 $h^{-1}$ cMpc. Additional lines of sight drawn along random directions through the simulation were spliced together until each sightline had a total length of 220 $h^{-1}$ cMpc. We selected haloes from four snapshots with volume-weighted average neutral fractions $\langle x_{\rm HI} \rangle_{\rm v}$ = (0.29, 0.47, 0.71, 0.91), i.e., separated by $\Delta \langle x_{\rm HI} \rangle_{\rm v} \sim 0.2$. This choice was made due to the simulation snapshots available to us, which were taken at fixed redshifts rather than specific volume-weighted average neutral fractions.  

As will be discussed in Section \ref{sec:shape}, the characteristic shape of these damping wings evolves with redshift. Unless otherwise specified, all the mock damping wings were generated from lines of sight with densities, distances and velocities rescaled to $z = 6.54$ as in \citet{chen2023}. The Ly$\alpha$ absorption was computed with contributions of gas in the foreground of the halo using the analytic approximation to the Voigt profile presented in \citet{teppergarcia2006}. For most of this Letter, we only include the contribution of gas with $x_{\rm HI}$ > 0.5 when computing the Ly$\alpha$ absorption, as in \citet{chen2023}. We are therefore neglecting any contribution from residual neutral hydrogen inside ionized bubbles. This assumption will be relaxed in Section \ref{sec:obs}. 

\section{Characteristic shape of IGM damping wings}
\label{sec:shape}

\begin{figure}
\begin{center}
	\includegraphics[width=0.75\columnwidth]{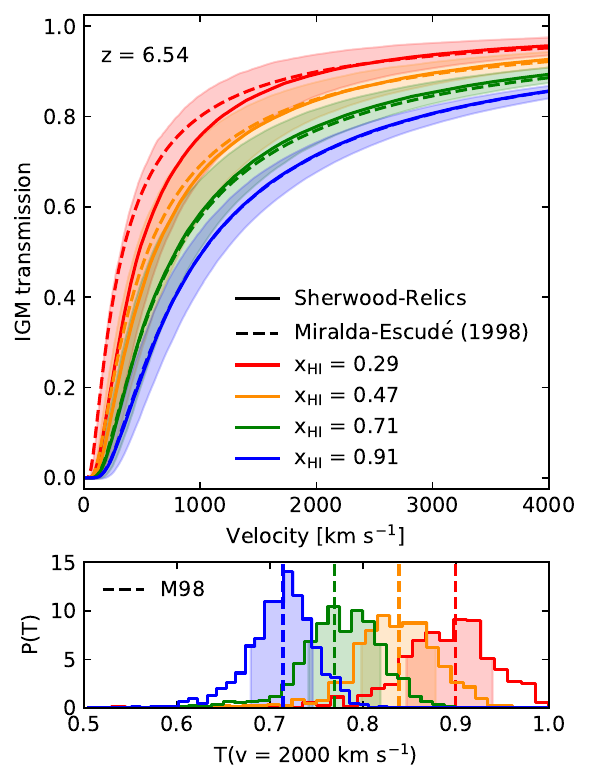}
    \caption{\textit{Top:} Mock IGM damping wings realigned along the velocity axis such that the different transmission curves all reach a transmission $T=0$ at the same point. The different colours represent different volume-weighted average neutral fractions. The solid lines are the median values of the transmission curves generated from the Sherwood-Relics patchy reionization simulation and the shaded regions show the 68 per cent scatter around the median. The dashed lines are the realigned transmission curves calculated using the \citet{miraldaescude1998} damping wing model, assuming the same volume-weighted average neutral fraction. \textit{Bottom:} Probability density function of the IGM transmission $T$ measured at a velocity of $2000$ km s$^{-1}$ (after realignment). The shaded region shows the 68 per cent scatter as plotted in the top panel. The vertical dashed lines mark the transmission calculated from the \citet{miraldaescude1998} model at the same velocity offset.}
    \label{fig:damping_wings}
    \end{center}
\end{figure}

\subsection{Shape at fixed redshift}
\label{sec:shape_fixz}

\begin{figure*}
	\begin{center}
	    
	\includegraphics[width=2\columnwidth]{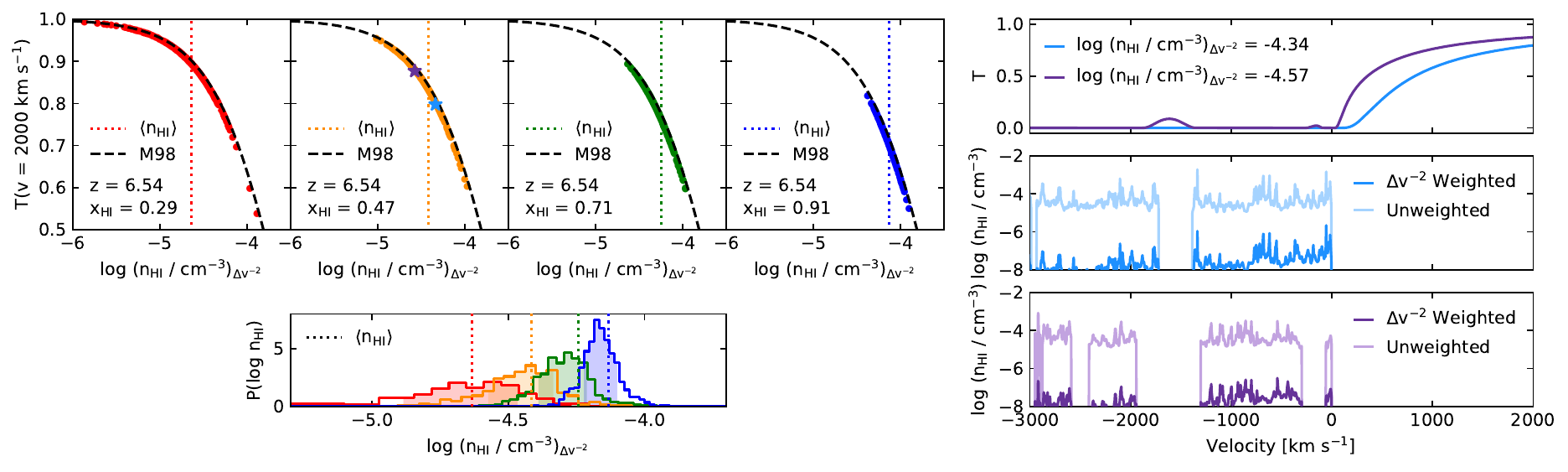}
    \caption{\textit{Top left:} The coloured points show the IGM transmission at $v=2000$ km s$^{-1}$ as a function of the average \ion{H}{i} number density along the line of sight, weighted by its contribution to the damping wing. Each panel shows results at a different volume-weighted average IGM neutral fraction. The vertical dotted line shows the corresponding mean \ion{H}{i} number density. The black dashed lines are the predictions from the \citet{miraldaescude1998} analytic model for the IGM damping wing and are the same in all panels. \textit{Bottom left:} Probability density function of the average \ion{H}{i} number density along a line of sight, weighted by $\Delta v^{-2}$ as above. The shaded regions show the 68 per cent scatter. We note that the distribution for the model with $\langle x_{\rm HI} \rangle_{\rm v} = 0.29$ has a long tail of average \ion{H}{i} number densities extending to below what is plotted here. These correspond to the most ionized sightlines. The vertical dotted line shows the mean \ion{H}{i} number density at each volume-weighted average IGM neutral fraction. \textit{Right:} The top panel shows weak (purple) and strong (blue) IGM transmission curves that have been realigned such that their transmission reaches $T=0$ at the same point, drawn from a simulation with $\langle x_{\rm HI} \rangle_{\rm v} = 0.47$ (the sightlines are indicated by the stars in that panel on the left). The light lines in the middle and bottom panels show the \ion{H}{i} number density along the line of sight, with the colours corresponding to the spectra in the top panel. The darker lines show the \ion{H}{i} number density, weighted by its contribution to the average value as plotted in the panels on the left.}
    \label{fig:damping_wings_density_fixz}
\end{center}
\end{figure*}

Using the mock IGM transmission spectra, we first demonstrate that we recover the characteristic damping wing shape as a function of volume-weighted average neutral fraction as presented in \citet{chen2023} (top panel of Figure \ref{fig:damping_wings}). We indeed find that, after the mock damping wings have been shifted along the velocity axis so that $T=0$ at the same point, the shapes of the damping wings generated at each volume-weighted average neutral fraction strongly resemble each other. We also plot the IGM damping wings calculated using the analytic model of \citet{miraldaescude1998}, which assumes that all of the gas is at the mean density at that redshift and that there is a single average value of $x_{\rm HI}$ everywhere. As in \citet{chen2023}, we find that these ``average'' damping wings are very similar to the patchy reionization model, after realignment. We find that there is more overlap in the 68 per cent scatter of our damping wing profiles at different $\langle x_{\rm HI} \rangle_{\rm v}$ than in \citet{chen2023}. This may be because we probe a slightly narrower range of models with $\Delta \langle x_{\rm HI} \rangle_{\rm v} \sim 0.2$, rather than $\Delta \langle x_{\rm HI} \rangle_{\rm v} \sim 0.25$ in that work. We further find that there is significant overlap between the IGM  transmission from the different $\langle x_{\rm HI} \rangle_{\rm v}$ outputs outside of the 68 per cent scatter (bottom panel of Figure \ref{fig:damping_wings}).

To quantify the source of this scatter, we consider the properties of the gas surrounding the host galaxies and how that may influence the shape of the damping wing. In particular, we investigate how the average \ion{H}{i} number density along the line of sight correlates with the IGM transmission. However, it is important to apply the correct weighting when doing the averaging. The Voigt profile is proportional to $v^{-2}$ in the Lorentzian wings of the profile, where $v$ is the velocity. We therefore determine the average \ion{H}{i} number density by weighting this quantity by $(\Delta v)^{-2}$. Here we have defined $\Delta v =  |v_{\rm pixel} - v_{\rm T}|$, where $v_{\rm pixel}$ is position of a given pixel on the velocity axis of our sightline and $v_{\rm T}$ is the velocity offset behind the realignment point at which we measure the IGM transmission. The results are shown in the left panels of Figure \ref{fig:damping_wings_density_fixz}, assuming $v_{\rm T} = 2000$ km s$^{-1}$. Our results are not sensitive to this choice of velocity offset.

We compare the results for our simulated sightlines to the \citet{miraldaescude1998} damping wing model for different values of the \ion{H}{i} number density. This model typically assumes a \ion{H}{i} number density $n_{\rm HI} = x_{\rm HI} \, \langle n_{\rm H} \rangle$, where $n_{\rm HI}$ is the \ion{H}{i} number density, $x_{\rm HI}$ is the average IGM neutral fraction and $\langle n_{\rm H} \rangle$ is the background hydrogen density. However, here we wish to explore how variations in the local $n_{\rm HI}$ impact the IGM transmission, so we explore a range of values. We find that the transmission is correlated with the average \ion{H}{i} number density along the sightline, with higher (lower) average \ion{H}{i} densities producing stronger (weaker) damping wings. The \citet{miraldaescude1998} model reproduces the relation between IGM transmission and average \ion{H}{i} number density very well.  

We find that the average \ion{H}{i} number density along our lines of sight has a substantial scatter, but is clustered around the volume-weighted average \ion{H}{i} number density for each choice of $\langle x_{\rm HI} \rangle_{\rm v}$. This explains the similarity between the transmission curves for the patchy simulation and \citet{miraldaescude1998} model in Figure \ref{fig:damping_wings}. The scatter is due to the location of galaxies within ionized bubbles, the clustering of these bubbles and the density of the environment in which the galaxy lives. An example of this is shown for two realigned sightlines in the right panels of Figure \ref{fig:damping_wings_density_fixz}. In both cases, there is an island of neutral hydrogen at the point where $T=0$. For the stronger damping wing, this marks the beginning of a long neutral island. However, for the weaker damping wing, this island marks only a small wall between two ionized bubbles, resulting in a lower average \ion{H}{i} number density. We also find that there is overlap in average \ion{H}{i} number densities measured along different lines of sight between the models with different volume-averaged IGM neutral fractions (bottom left of Figure \ref{fig:damping_wings_density_fixz}). This explains the overlap in the scatter of the characteristic damping wing shapes at different points in the reionization history.   

\begin{figure}
	\begin{center}
	\includegraphics[width=0.75\columnwidth]{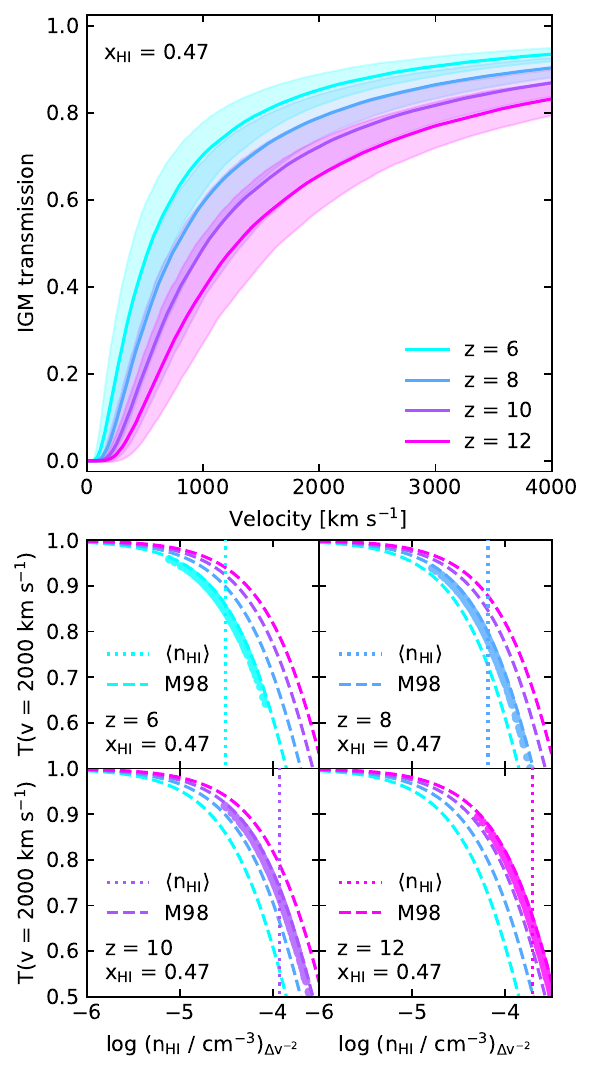}
    \caption{\textit{Top:} Mock IGM damping wings realigned along the velocity axis such that the different transmission curves all reach a transmission $T=0$ at the same point. The different colours represent different redshifts. The solid lines are the median values of the transmission curves and the shaded regions show the 68 per cent scatter around the median. \textit{Bottom:} The points show the IGM transmission at $2000$ km s$^{-1}$ as a function of the average \ion{H}{i} number density along the line of sight, weighted by its contribution to the damping wing. Each panel shows results at a different redshift. The vertical dotted line shows the mean \ion{H}{i} number density at each redshift. The dashed lines are the predictions from the \citet{miraldaescude1998} analytic model for the IGM damping wing, which evolves with redshift.}
    \label{fig:damping_wings_density_fixxHI}
\end{center}
\end{figure}

\subsection{Shape at fixed volume-weighted average neutral fraction}

We next investigate how the characteristic damping wing shape evolves with redshift, holding the volume-weighted average neutral fraction fixed at $\langle x_{\rm HI} \rangle_{\rm v} = 0.47$. We rescale the sightlines from this simulation output to redshifts $z=6,8,10$ and 12. The resulting realigned transmission curves are shown in the top panel of Figure \ref{fig:damping_wings_density_fixxHI}. We find that the damping wings become stronger towards higher redshift. Part of this is because the \ion{H}{i} number density along a line of sight will increase as $n_{\rm HI} \propto (1 + z)^3$. As shown in the bottom panels of Figure \ref{fig:damping_wings_density_fixxHI}, we do find that the densities are higher as we move to higher redshift, both in the mean values of the cosmological background density at that redshift and in the line of sight averages.

However, this does not completely explain all of the evolution. We find that even at fixed proper \ion{H}{i} number density, we see stronger damping wings at higher redshifts. This is because of additional effects due to the expansion of the Universe when calculating the optical depth, such as the conversion from proper length to redshift. We plot the expected transmission in the \citet{miraldaescude1998} model for the four redshift bins we consider as the four dashed lines in each panel of Figure \ref{fig:damping_wings_density_fixxHI} (where we have plotted the lines for all redshifts in each panel for ease of comparison, but note that the points only correspond to one line in each panel). Again, the analytic model nicely recovers the trend found in the simulated sightlines.

\section{Prospects for observations}
\label{sec:obs}

\begin{figure*}
	\begin{center}
	\includegraphics[width=1.85\columnwidth]{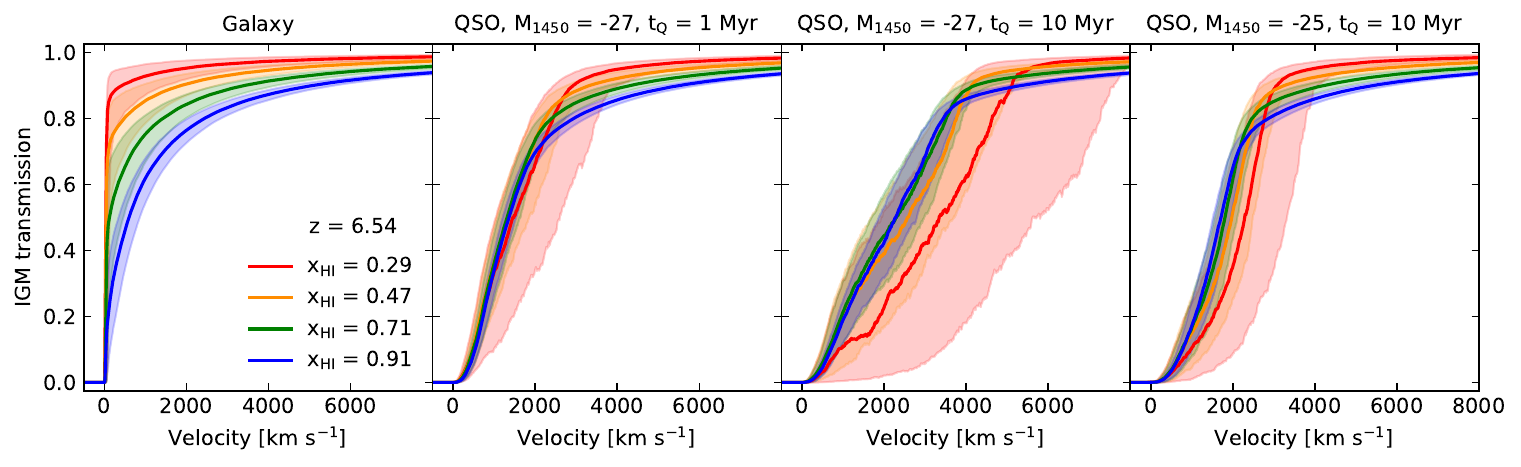}
    \caption{Mock IGM damping wings realigned along the velocity axis such that the different transmission curves all reach a transmission $T=10^{-6}$ at the same point. The colours represent different volume-weighted average IGM neutral fractions. The solid line is the median transmission curve and the shaded regions show the 68 per cent scatter around the median. Note that we plot a wider velocity range here than in previous figures. \textit{Left:} Transmission curves generated from a Sherwood-Relics patchy reionization simulation including the residual neutral hydrogen in the bubbles. These damping wings are what would be expected for galaxies or gamma-ray burst afterglows, ignoring any intrinsic absorption. \textit{Middle left:} Transmission curves generated from a Sherwood-Relics patchy reionization simulation also assuming ionization by a quasar with $M_{1450}=-27$ that has been shining for 1 Myr. \textit{Middle right:} As before, but for a quasar that has been shining for 10 Myr. \textit{Right:} As before, but for a fainter quasar with $M_{1450}=-25$.}
    \label{fig:damping_wings_gal_qso}
\end{center}
\end{figure*}

So far we have only presented idealised results, because we have assumed that there is no residual neutral gas inside the ionized bubbles (following \citealt{chen2023}). However, even a small amount of residual neutral hydrogen can saturate the absorption in the Ly$\alpha$ forest, at neutral fractions as low as $x_{\rm HI} \sim 10^{-5}$ \citep{mcquinn2016}. We have also thus far neglected the peculiar velocities of infalling gas which can impact Ly$\alpha$ transmission, but this will be a small effect across the scales of several thousand km s$^{-1}$ that we consider here. We now investigate how absorption by the residual neutral hydrogen impacts the observability of the characteristic shape of damping wings.

We first calculate the Ly$\alpha$ optical depth for the same set of sightlines as analysed in Figure \ref{fig:damping_wings}, but now including the effects of all the neutral gas, both in the neutral islands and within the ionized bubbles, as well as the peculiar velocity of the gas. These transmission curves are representative of the IGM damping wings that may be seen in high-redshift galaxies or gamma-ray burst afterglows, where the sources live within ionized bubbles carved out by galaxies. We repeat the process of realigning the transmission curves. However, due to small amounts of transmitted flux in the partially-ionized regions, we now realign the curves such that they all reach $T=10^{-6}$ at the same point. The results are shown in the left panel of Figure \ref{fig:damping_wings_gal_qso}. We find that in contrast to what was shown in \citet{chen2023} and Figure \ref{fig:damping_wings}, we no longer see a smooth decline in the IGM transmission with decreasing velocity, but rather a sharp vertical cutoff in the transmission. This arises from the residual neutral hydrogen in the ionized bubbles \citep[e.g.,][]{mason2020}. Despite this, we still find that the median of the models with different volume-weighted average IGM neutral fractions can be distinguished at large velocity offsets from the realignment point. This may therefore be a useful way to probe reionization in a large sample of galaxies that JWST will discover, stacked in different redshift bins. However, noise in the observations will make it challenging to find the point where the transmission first reaches $T \approx 0$. It will also be necessary to accurately model the intrinsic spectrum of the galaxies to recover the normalised IGM transmission. Finally, absorption within the galaxy itself will further confuse the signal \citep{heintz2023}.

It is also useful to consider how this technique may apply to damping wings in high-redshift quasars. Currently, only a handful of quasars are known above redshift 7 \citep{fan2022}. However, with the advent of Euclid, many fainter quasars are predicted to be discovered \citep{schindler2023} which may be good candidates for stacking at different redshifts and searching for these characteristic damping wing shapes. To account for the extra ionization by the quasar of its surroundings, we post-process the sightlines with the 1D radiative transfer code presented in \citet{bolton2007}. We assume a magnitude $M_{1450} = -27$, similar to the luminosity of the brightest known quasars. We assume the spectrum is a broken power law, with $f_{\nu} \propto \nu^{-0.5}$ for 1050 \AA $\, < \lambda < 1450$ \AA \, and $f_{\nu} \propto \nu^{-1.5}$ for $\lambda < 1050$ \AA. This results in an ionizing photon rate $\dot{N}_{\gamma} = 1.9 \times 10^{57}$ s$^{-1}$. Euclid is expected to discover many fainter quasars, so we also investigate a quasar with $M_{1450} = -25$ ($\dot{N}_{\gamma} = 3.0 \times 10^{56}$ s$^{-1}$). We show here results for two quasar lifetimes, $t_{\rm Q}$ = 1 Myr and 10 Myr. However, in reality, the quasars will sample a range of magnitudes and lifetimes \citep[e.g.,][]{eilers2017}.

Stacking the quasar transmission spectra is more complicated than for the previous case, as absorption within the quasar proximity zone can reach $T \approx 0$, even though there may be more transmission towards lower redshifts. To account for this, we smooth the transmission spectra with a boxcar filter with width 20 \AA \, in the observed frame, as is done for measuring quasar proximity zone sizes \citep{fan2006}. We then realign the transmission curves based on where this smoothed spectrum reaches $T=10^{-6}$ for the first time, and stack the smoothed curves to calculate the median and scatter of the transmission. The results are shown in Figure \ref{fig:damping_wings_gal_qso}. We find that for a magnitude $M_{1450} = -27$ and a lifetime of 1 Myr, we do not see a sharp cutoff in the transmission as was seen for the galaxy transmission curves. This is because the ionizing photons from the quasar have ionized much of the residual neutral hydrogen in the bubbles. At large velocity offset ($v \gtrsim 3000$ km s$^{-1}$), we find that it is possible to differentiate between the median of the stacked transmission curves at different values of $\langle x_{\rm HI} \rangle_{\rm v}$. At lower velocity offset, it is more difficult as there is a larger scatter in transmission at low $\langle x_{\rm HI} \rangle_{\rm v}$. This is due to the scatter in proximity zone sizes becoming larger as the neutral islands which impede the quasar ionization front become rarer.

For a longer quasar lifetime, even though the quasar has had more time to ionize its surroundings, there again remains a trend with the volume-weighted average IGM neutral fraction at large velocity offsets and high $\langle x_{\rm HI} \rangle_{\rm v}$. We find that the damping wings of fainter quasars at longer lifetimes look more like those of young bright ones, and may therefore be a useful population on which to apply this technique. It may also be possible to improve the method by stacking quasars with similar proximity zone sizes, or choosing a different value of the IGM transmission at which to align the spectra. However, as with the galaxy damping wings, the noise of the observations and intrinsic spectrum of the quasar will make the analysis more complex.

\section{Conclusions}
\label{sec:conclusions}

We have presented a closer look at the characteristic shape of intergalactic damping wings during reionization that was first identified in \citet{chen2023}. Using a simulation of inhomogeneous reionization from the Sherwood-Relics simulation suite, we found that we can recover the same trend presented in that work. Once the transmission curves are realigned along the velocity axis so that they reach $T=0$ at the same point, the median transmission curve is a strong function of volume-weighted average IGM neutral fraction at fixed redshift, but there is overlap between the scatter of realigned transmission curves generated at different points in the reionization history. 

We demonstrated that this characteristic shape arises because the damping wings are set by the average \ion{H}{i} number density along the line of sight, weighted by its contribution to the optical depth. Scatter in this quantity is driven by fluctuations in the density and ionization field, and higher (lower) average \ion{H}{i} number densities result in stronger (weaker) IGM damping wings. We also found that the characteristic shape of these damping wings evolves with redshift, due to the expansion of the Universe. These effects are all well captured by the \citet{miraldaescude1998} IGM damping wing model.

We finished by showing that although more realistic models of damping wings in galaxies and quasars distort the characteristic damping wing shape, it is still possible to tell the difference between transmission curves drawn from simulations with different $\langle x_{\rm HI} \rangle_{\rm v}$ at large velocity offset from the realignment point. This suggests that realigning IGM transmission curves may be a promising avenue for constraining reionization with large samples of faint objects delivered by JWST and Euclid. 

\section*{Acknowledgements}

The simulations used in this work were performed using the Joliot Curie supercomputer at the Tr{\'e}s Grand Centre de Calcul (TGCC) and the Cambridge Service for Data Driven Discovery (CSD3), part of which is operated by the University of Cambridge Research Computing on behalf of the STFC DiRAC HPC Facility (www.dirac.ac.uk).  We acknowledge the Partnership for Advanced Computing in Europe (PRACE) for awarding us time on Joliot Curie in the 16th call. The DiRAC component of CSD3 was funded by BEIS capital funding via STFC capital grants ST/P002307/1 and ST/R002452/1 and STFC operations grant ST/R00689X/1.  This work also used the DiRAC@Durham facility managed by the Institute for Computational Cosmology on behalf of the STFC DiRAC HPC Facility. The equipment was funded by BEIS capital funding via STFC capital grants ST/P002293/1 and ST/R002371/1, Durham University and STFC operations grant ST/R000832/1. DiRAC is part of the National e-Infrastructure.  JSB is supported by STFC consolidated grant ST/T000171/1. MGH is supported by STFC consolidated grant ST/S000623/1. GK is partly supported by the Department of Atomic Energy (Government of India) research project with Project Identification Number RTI~4002, and by the Max Planck Society through a Max Planck Partner Group. Support by ERC Advanced Grant 320596 ‘The Emergence of Structure During the Epoch of Reionization’ is gratefully acknowledged. We thank Volker Springel for making \textsc{p-gadget-3} available. We also thank Dominique Aubert for sharing the \textsc{aton} code. For the purpose of open access, the author has applied a Creative Commons Attribution (CC BY) licence to any Author Accepted Manuscript version arising from this submission.

\section*{Data Availability}

All data and analysis code used in this work are available from the first author on reasonable request.



\bibliographystyle{mnras}
\bibliography{ref} 



\bsp	
\label{lastpage}
\end{document}